# Orbital design of topological insulators from two-dimensional semiconductors

*Lei Gao*[1, #], *Jia-Tao Sun*[1,3, #], *Gurjyot Sethi*[2], *Yu-Yang Zhang*[1, 4], *Shixuan Du*[1, 4, †], *Feng Liu*[2, ‡]

1. Institute of Physics & University of Chinese Academy of Sciences, Chinese Academy of Sciences, Beijing 100190, P.R. China
2. Department of Materials Science and Engineering, University of Utah, Salt Lake City, Utah 84112, USA
3. School of Information and Electronics, Beijing Institute of Technology, Beijing 100081, China
4. CAS Center for Excellence in Topological Quantum Computation, Beijing 100190, P.R. China

[#] Contributed equally
[†‡] Corresponding author. E-mail: [†]sxdu@iphy.ac.cn; [‡]fliu@eng.utah.edu

Two-dimensional (2D) materials have attracted much recent attention because they exhibit various distinct intrinsic properties/functionalities, which are, however, usually not interchangeable. Interestingly, here we propose a generic approach to convert 2D semiconductors, which are amply abundant, to 2D topological insulators (TIs), which are less available, via selective atomic adsorption and strain engineering. The approach is underlined by an orbital design principle that involves introducing an extrinsic *s*-orbital state into the intrinsic *sp*-bands of a 2D semiconductor, so as to induce *s-p* band inversion for a TI phase, as demonstrated by tight-binding model analyses. Remarkably, based on first-principles calculations, we apply this approach to convert the semiconducting monolayer CuS and CuTe into a TI by adsorbing Na and K respectively with a proper *s*-level energy, and CuSe into a TI by adsorbing a mixture of Na and K with a tuned *s*-level energy or by adsorbing either Na or K on a strained CuSe with a tuned *p*-level valence band edge. Our findings open a new door to the discovery of TIs by a predictive materials design, beyond finding a preexisting 2D TI.

Recently, two-dimensional (2D) materials have attracted much attention for both fundamental interest and their potential applicaitions [1-4]. So far, a suite of different classes of 2D materials have been discovered, ranging, e.g., from inorganic Dirac semimetal [5, 6], semiconductor [7, 8], topological insulator (TIs) [9-11], superconductor [12] and ferromagnet [13], to their organic counterparts [14-17]. Although some efforts have also been made to tune the intrinsic properties of 2D materials, such as by strain [18] and alloying [19], usually different classes of 2D materials are not interchangeable. In this Letter, we propose a generic approach to convert 2D semiconductors, which are amply abundant, to 2D TIs, which are relatively less available, via surface adsorption and strain engineering.

Becasuse of an infinitely large surface-to-volume ratio, 2D materials are extremely sensitive to surface modification [20, 21]; while their unltrathinness makes them very sensitive to strain [21-23]. This offers them a unique advantage over their 3D counterparts, as the intrinsic properties of the whole 2D materials can in principle be modified through adsorption and/or strain engineering so that one class of 2D materials may be converted into another class. Here we demonstrate this possibility by converting 2D semiconductors into TIs predominantly by surface adsorption where one alkali atom is introduced into each unit cell of the 2D semiconductor to completely change its intrinsic band structure and band topology, which is apparently impossible for 3D materials.

We will first illustrate our approach, which is underlined by an orbital design principle, using a tight-binding (TB) model [24] that represents a typical 2D $sp$-bands semiconductor in a trigonal lattice. Their top of the valence band and bottom of the conduction band mainly consists of $p$- and $s$-state, respectively. An extrinsic $s$-state of alkali metal with appropriate energy is brought in to induce $s$-$p$ band inversion to form a TI phase; a generic phase diagram is constructed in the parameter space of energy levels and spin-orbit coupling (SOC). Then, using first-principles calculations, we demonstrate this approach for real materials by converting the semiconducting monolayer CuS and CuTe into TI via adsorbing Na and K respectively with a proper $s$-level energy, and CuSe into TI via adsorbing a mixture of Na and K with a tuned $s$-level energy or by adsorbing either Na or K on a strained CuSe via a tuned $p$-level valence-band edge.

We prescribe this concept of orbital design principle in a three-site (alkali metal adsorbate, cation and anion in a compound semiconductor) and four-band ($s_1^A$, $s_2^S$, $p_x^S$ and $p_y^S$) TB model. Expending to the first-order of $k$ around the $\Gamma$ point, the spinless Hamiltonian reduces to

$$H = \begin{pmatrix} \varepsilon_{s_1} + 6t_{s_1 s_1 \sigma} & 3t_{s_1 s_2 \sigma} & -\frac{\sqrt{3}}{4}t_{s_1 p\sigma}(ik_x + k_y) & -\frac{\sqrt{3}}{4}t_{s_1 p\sigma}(ik_x - k_y) \\ & \varepsilon_{s_2} + 6t_{s_2 s_2 \sigma} & -\frac{\sqrt{3}}{2}t_{s_2 p\sigma}(ik_x + k_y) & -\frac{\sqrt{3}}{2}t_{s_2 p\sigma}(ik_x - k_y) \\ & & \varepsilon_p + 3(t_{pp\sigma} + t_{pp\pi}) + \lambda & 0 \\ & & & \varepsilon_p + 3(t_{pp\sigma} + t_{pp\pi}) - \lambda \end{pmatrix}, \quad (7)$$

where $\varepsilon_{s_1}$, $\varepsilon_{s_2}$ and $\varepsilon_p$ are the on-site energies; $t_{s_1 s_1 \sigma}$, $t_{s_1 s_2 \sigma}$, $t_{s_2 s_2 \sigma}$, $t_{s_1 p\sigma}$, $t_{s_2 p\sigma}$, $t_{pp\sigma}$ and $t_{pp\pi}$ are nearest-neighbor (NN) hopping parameters; $\lambda$ is SOC strength; $k_x$ and $k_y$ are momenta along $x$ and $y$ direction respectively. The whole TB Hamiltonian is available in Supplemental Material.

At the $\Gamma$ point, the four eigenvalues are $E_{s_1} = \varepsilon_{s_1} + 6t_{s_1 s_1 \sigma}$, $E_{s_2} = \varepsilon_{s_2} + 6t_{s_2 s_2 \sigma}$ and $E_p^{\pm\lambda} = E_p \pm \lambda = \varepsilon_p + 3(t_{pp\sigma} + t_{pp\pi}) \pm \lambda$, respectively. $E_{s_1}$ and $E_{s_2}$ are independent of SOC, while $E_p$ are split by SOC with $\Delta E_p^{SOC} = 2\lambda$. As shown in Fig. 1(a), when the extrinsic $s_1$ band with an appropriate energy is brought into the proximity of intrinsic $s_2 p$-bands of a 2D semiconductor, an $s_1$-$p$ band inversion can happen, resulting in topological phase transition to form the TI phase. Depending on whether the $s_1$ band inverts with the upper or lower $p_x/p_y$ band, there can be two types of TI phases as indicated by TI(A) and TI(B) in the right panel of Fig. 1(a).

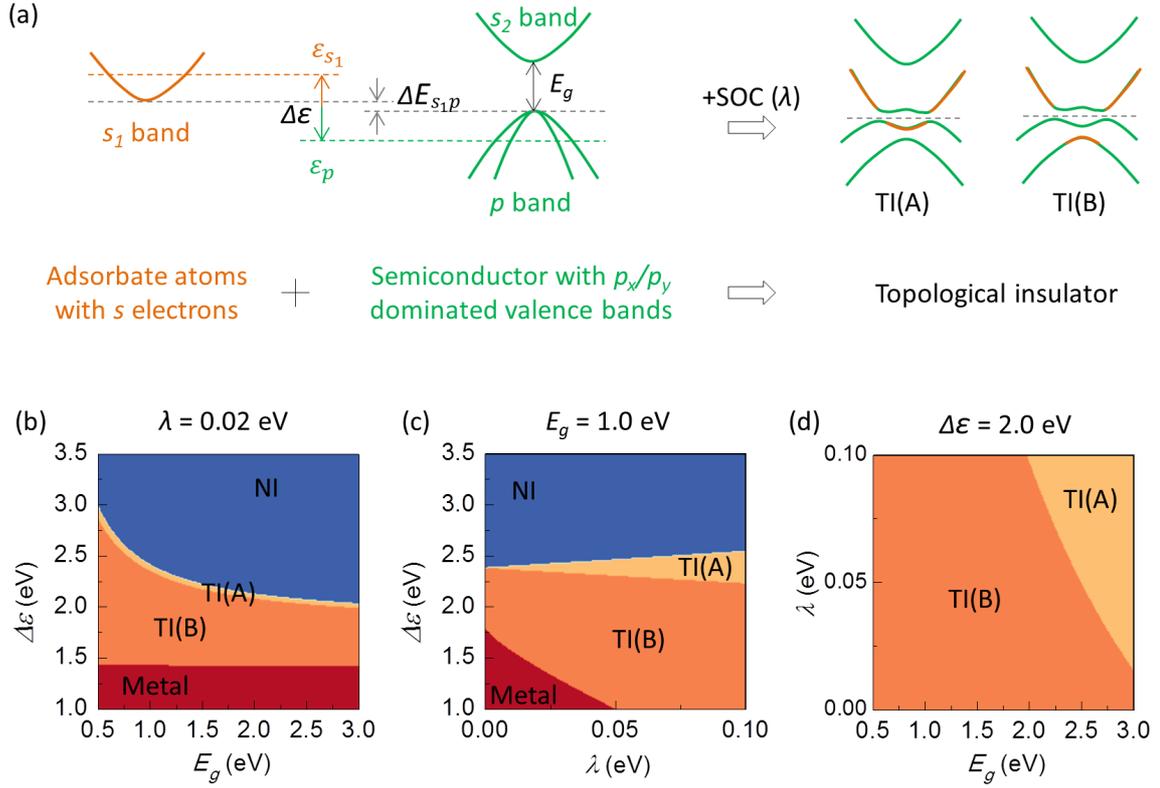

FIG 1. (a) Schematic illustration of orbital designed TIs from 2D semiconductors. The orange and green solid curves indicate the bands of the adsorbate atom and semiconductor, respectively. The dashed lines in the right panel indicate the Fermi level. TI(A) and TI(B) are defined for the $s_1$ band inverting with the upper and lower $p_x/p_y$ band, respectively. (b)-(d) Topological phase diagrams in the parameter space of $\Delta\varepsilon$ and $E_g$, $\Delta\varepsilon$ and $\lambda$, $\lambda$ and $E_g$, respectively. The other parameters are same: $t_{s_1s_1\sigma} = t_{s_1s_2\sigma} = t_{s_2s_2\sigma} = -0.25$ eV, $t_{s_1p\sigma} = -0.5$ eV, $t_{s_2p\sigma} = -0.1$ eV, $t_{pp\sigma} = 0.1$ eV, $t_{pp\pi} = 0.01$ eV and $\theta = \pi/3$.

To reveal how the orbital design principle works, we construct phase diagrams [Figs. 1(b)-1(d) and S2-S4] of electronic structure in the parameter of $\Delta\varepsilon = \varepsilon_{s_1} - \varepsilon_p$ (on-site energy difference), $E_g$ (band gap of semiconductor without SOC) and $\lambda$, based on the above TB model. With an increasing $\Delta\varepsilon$ at given $\lambda$ and $E_g$ [Figs. 1(b) and 1(c)], the phase evolves from a metal, TI(B), TI(A), to a normal insulator (NI). The typical band structure of each phase is shown in Fig. S1. The $s_1$-$p$ band inversion may or may not exist in the absence of SOC. When SOC is included, the degenerate $p_x/p_y$ valance bands split. As illustrated in Fig. 1(d), this may result in a TI phase with an appropriate $\Delta\varepsilon$, which can be either a TI(A) or TI(B) phase. The condition for the occurrence of TI (A) or TI(B) phase is determined by relative magnitude of band splitting $\Delta E_{s_1p} = E_{s_1} - E_p = \Delta\varepsilon - T$ (the $s_1/p$ energy difference at the Γ point without considering SOC as indicated in Fig. 1(a); $T$ is related to hopping parameters) and SOC gap $\Delta E_p^{SOC}$ [3, 24]. The TI phase can only occur for $\Delta E_{s_1p} < \Delta E_p^{SOC}$, otherwise only NI exists. Furthermore, for $\Delta E_{s_1p} < \Delta E_p^{SOC}$, only TI(A) occurs for $\Delta E_{s_1p} > 0$, where the SOC induces an $s_1$-$p$ band inversion similar to the case of small-gap quantum well [10]; for $\Delta E_{s_1p} < 0$, where the $s_1$-$p$ bands are already inverted and the SOC simply opens a gap similar to graphene [9], as determined by whether $|\Delta E_{s_1p}|$ is smaller or bigger than $\Delta E_p^{SOC}$, either TI(A) or TI(B) occurs. Apparently, besides $\Delta\varepsilon$, the smaller $E_g$ [Figs. 1(b) and S3] or bigger $\lambda$ [Figs. 1(c) and S2] is, the larger the range of TI phase will be.

Given the understanding of the orbital design principle based on TB model, it is natural to ask whether it can be realized in real materials? Recent experimental fabrication of monolayer CuSe [25, 26], which posseses the $p_x/p_y$ dominated valance bands, has drawn our attention. The monolayer honeycomb monochalcognides MX (M = Cu, Ag; X = S, Se, Te) family can all be viewed effectively as "hole-doped" $sp$-band 2D semiconductors, with one electron depleted per unit cell as shown in Fig. S5. It is expected that adsorption of one alkali atom per unit cell, which introduces one electron, will move the Fermi level up to make a semiconductor. Furthermore, if the energy of the $s_1$-state of alkali atom is right relative to the top of the valence $p$-band, the condition for the TI phase underlined by the above design principle can be satisfied.

Therefore, the key is to identify the right alkali element with an appropriate $s_1$-state energy for a given 2D semiconductor, such as CuSe with the known valence $p$-band position and SOC strength. Because of the level repulsion, the relative position of final $s_1$ and $p$ bands is also affected by the bottom of the conduction $s_2$ band. So, based on the above TB model, we construct another phase diagram for the appropriate $\varepsilon_{s_1}$ in the parameter space of $\varepsilon_{s_2}$ and $\varepsilon_p$, as shown in Fig. 2(a). It is found that the appropriate $\varepsilon_{s_1}$ of the adsorbate increases, when the $\varepsilon_{s_2}$ ($\varepsilon_p$) of the cation (anion) in the compound semiconductor increases (decreases). Note that an increased $\varepsilon_{s_1}$ and $\varepsilon_{s_2}$ means also an increased ability of donating electrons, while a decreased $\varepsilon_p$ means an increasing ability of accepting electrons.

Using the phase diagram Fig. 2(a) as a guide, we have identified the appropriate alkali elements as adsorbates to successfully convert several 2D semiconductors into TIs. Taking CuS, CuSe and CuTe as examples, which have the same cation but different anions, the ability of accepting electrons decreases from S, Se to Te, so that their $\varepsilon_p$ increases. Correspondingly, to convert them into TIs, the appropriate $\varepsilon_{s_1}$ to be chosen should decrease, with the adsorbate having a decreasing ability of donating electrons.

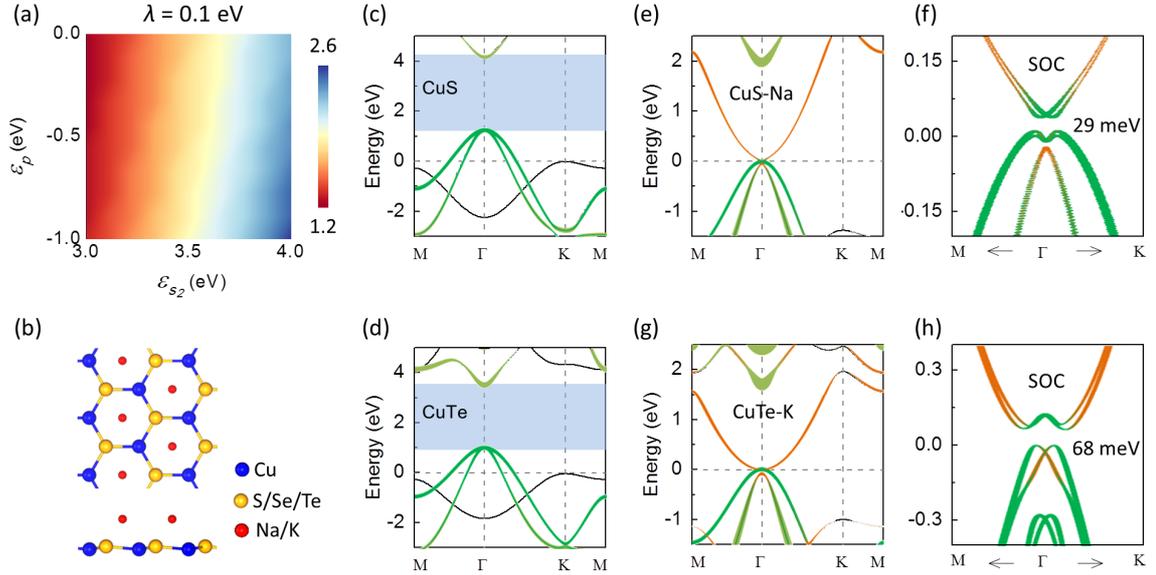

FIG 2. (a) Phase diagram of appropriate $\varepsilon_{s_1}$ in the parameter space of $\varepsilon_{s_2}$ and $\varepsilon_p$. The color indicates the value of $\varepsilon_{s_1}$. (b) Top and side views of monolayer MX-Y (M = Cu; X = S, Se, Te; Y = Na, K). (c)-(f) Projected band structures without SOC of CuS, CuTe, CuS-Na and CuTe-K, respectively. (g) and (h) Projected band structures with SOC of CuS-Na and CuTe-K, respectively. The width of colored lines represents the contribution from different orbitals. The components of the X-$s$ (X = Na, K) states are multiplied by a factor of 4 for clarity. The numbers indicate the band gap.

To comfirm the above hypothesis, first-principles calculations have been performed for CuS, CuSe and CuTe with different alkali adsorbates. Alkali atom prefers to be adsorbed at the center of hexagon as shown in Fig. 2(b). The "*p*-type" semiconductors CuS and CuTe [Figs. 2(c) and 2(d)] have been successfully converted into TIs [Figs. 2(e)-2(h)] via Na and K adsorption, respectively. Without considering SOC, the two degenerate $p_x/p_y$ bands are exactly at the Fermi level. And the $s_1$ band of adsorbate is below the $p_x/p_y$ bands of host semiconductor near the Γ point, leading to "pre-exsiting" band inversion. When SOC is included, the degenerate $p_x/p_y$ bands split with $\Delta E_p^{SOC}$, forming TI(B) and TI(A) phase for CuS-Na and CuTe-K, respectively. With the anion changing from S to Te, the appropriate adsorbate changes from Na to K with a decreasing ability of donating electrons, in good agreement with the design principle.

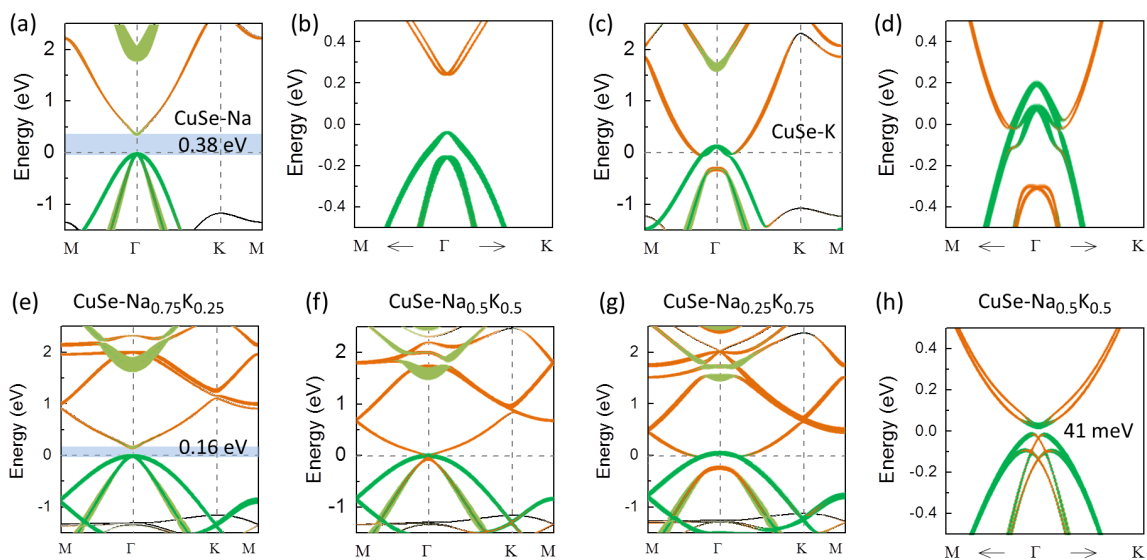

FIG 3. (a)-(d) Projected band structures of CuSe-Na and CuSe-K without and with SOC, respectively. (e)-(g) Projected band structures without SOC of CuSe-Na$_{0.75}$K$_{0.25}$, CuSe-Na$_{0.5}$K$_{0.5}$ and CuSe-Na$_{0.25}$K$_{0.75}$, respectively. (h) Projected band structures with SOC of CuSe-Na$_{0.5}$K$_{0.5}$. The width of colored lines represents the contribution from different orbitals. The components of the X-*s* (X = Na, K) states are multiplied by a factor of 4 for clarity. The numbers indicate the band gap.

Next, we further investigate the band structures of CuSe-Na [Figs. 3(a) and 3(b)] and CuSe-K [Figs. 3(c) and 3(d)], which are NI and metal respectively. Because the ability of donating electrons for Se is between that for S and Te, the $\varepsilon_{s_1}$ of Na and K is too big and too small, respectively, to convert CuSe into TI. Thus, according to the phase diagram [Fig. 2(a)], the appropriate $\varepsilon_{s_1}$ is between that of Na and K, indicating that adsorbing a mixture of Na and K with a tuned $s_1$-level energy may convert CuSe into TI. This is indeed confirmed by calculations of CuSe via adsorbing mixtures of Na and K with different ratios as shown in Figs. 3(e)-3(h). With the Na/K ratio decreasing, the $s_1$ band moves downward at the Γ point and the band structure evolves from CuSe-Na to CuSe-K with the band inversion occurring at CuSe-Na$_{0.5}$K$_{0.5}$. When SOC is included, CuSe-Na$_{0.5}$K$_{0.5}$ is found a TI possessing a non-trivial gap of 41 meV as shown in Fig. 3(h).

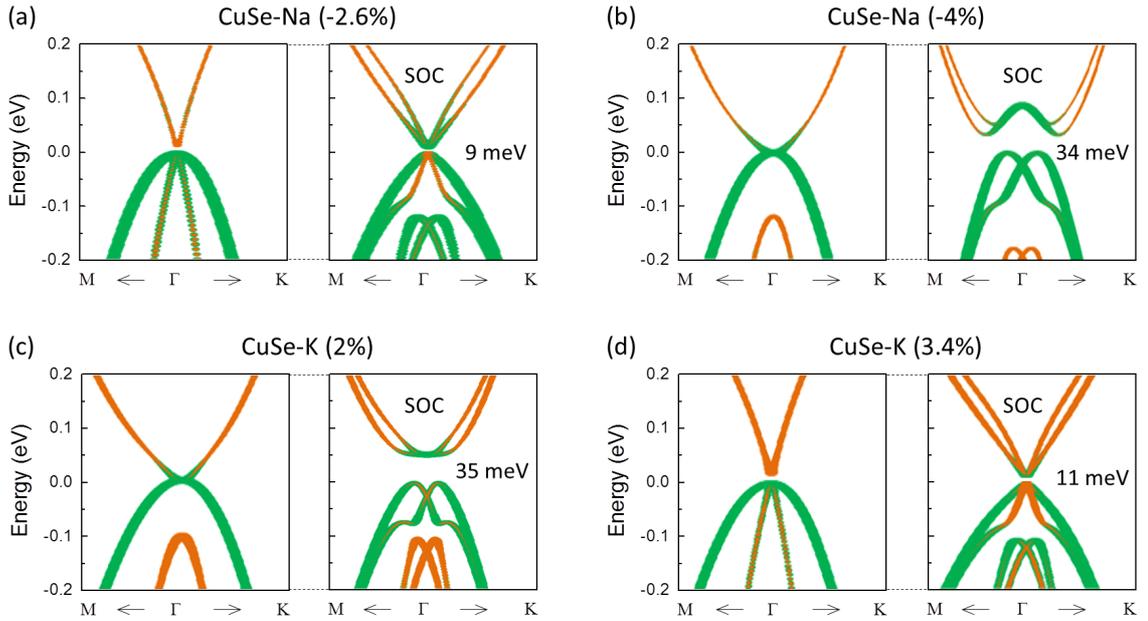

FIG 4. (a) and (b) Projected band structures of CuSe-Na with -2.6% and -4% compressive strain, respectively. (c) and (d) Projected band structures of CuSe-K with 2% and 3.4% tensile strain, respectively. The width of colored lines represents the contribution from different orbitals. The components of the X-$s$ (X= Na, K) states are multiplied by a factor of 4 for clarity. The numbers indicate the band gap.

Besides tuning the $s_1$ band via adsorbing different alkali elements, one can also slightly tune the $p_x/p_y$ valence bands of host semiconductor via strain so as to achieve a desired $\Delta E_{s_1 p}$. For example, CuSe-Na is a NI. By applying a compressive strain, the relative energy of $s_1$ band to the $p_x/p_y$ bands decreases near the Γ point [Figs. S6]. Then, the phase evolves from a NI, TI(A), TI(B) to a metal with the increasing strain [Figs. S6]. The specific band structure of TI(A) and TI(B) phases are shown in Figs. 4(a) and 4(b), respectively, for CuSe-Na with -2.6% and -4% compressive strain. To convert CuSe-Na into TI, the range of compressive strain is about -2.3% ~ -4.5%.

Similarly, CuSe-K is a metal. By applying a tensile strain, the relative energy of $s_1$ band to the $p_x/p_y$ bands increases near the Γ point [Figs. S7]. The phase evolves from a metal, TI(B), TI(A) to a NI with the increasing strain [Figs. S7]. The specific band structure of TI(B) and TI(A) phases are shown in Figs. 4(c) and 4(d), respectively, for CuSe-K with 2% and 3.4% tensile strain. To convert CuSe-K into TI, the range of tensile strain is about 1% ~ 3.6%.

In conclusion, we propose a novel approach underlined by an orbital design principle, to rationally convert 2D $sp$-band semiconductors into TIs. We envision this approach is generally applicable to a large number of 2D material databases. As prototypical examples, the feasibility of the approach is comfirmed by first-principles calculations of monolayer CuX (X = S, Se, Te) via adsorbing alkali atoms. Deponding on the desired relative position of extrinsic $s$-band of alkali atoms to the intrinsic top of $p$-valence band of host, a mixture of alkali atoms and either tensile or compressive strains may be needed. Our findings not only enrich 2D TIs with new material classes but also revive the discovery of TIs by predictive material disigns.

ACKNOWLEDGEMENTS

This work was financially supported by National Key Research and Development Projects of China (2016YFA0202300), Strategic Priority Research Program of Chinese Academy of Sciences (Grant Nos. XDB30000000 and XDB07030100), the National Natural Science Foundation of China (No. 51872284 and 61888102), the International Partnership Program of Chinese Academy of Sciences (No. 112111KYSB20160061), and the CAS Pioneer Hundred Talents Program, Beijing Nova Program (No. Z181100006218023). This project was also supported by the CAS Key Laboratory of Vacuum Physics and the Beijing Key Laboratory for Nanomaterials and Nanodevices. G.S. and F.L. at Utah acknowledge the support from US-DOE (Grant No. DE-FG02-04ER46148).